\documentclass[%
 aip,
cp,  
 amsmath,amssymb,
 reprint,%
]{revtex4-2}

\usepackage{graphicx}
\usepackage{dcolumn}
\usepackage{bm}

\usepackage[utf8]{inputenc}
\usepackage[T1]{fontenc}
\usepackage{mathptmx} 
\usepackage{indentfirst}

\begin{document}

\title{Software Suite for Modeling Cherenkov Images of Extensive Air Showers in the SPHERE-3 Detector}

\author{V.A. Ivanov} 
\email[Corresponding author: ]{ivanov.va18@physics.msu.ru}
\author{V.I. Galkin}
\author{E.A. Bonvech}
\author{T.M. Roganova}
\author{D.V. Chernov}
\author{D.A. Podgrudkov}
\author{M.D. Ziva}
\author{V.S. Latypova}
\author{C. Azra}
\author{E.L. Entina}
 \affiliation{Skobeltsyn Institute of Nuclear Physics, Moscow State University, Moscow, Russia.}

\date{\today} 

\begin{abstract}
The study of cosmic rays in the energy range from 1 to 1000 PeV is crucial for understanding their origins and propagation paths. As part of this research, a new SPHERE-3 installation is being developed, featuring enhanced light sensitivity and optical resolution, based on the experience gained with the balloon-borne SPHERE-2 installation. This report describes a computational complex designed for simulating the formation of Cherenkov light on the detector grid of the SPHERE-3 telescope. 

\vspace{\baselineskip}

Published as a conference paper at X International Conference on Mathematical Modeling.
\end{abstract}

\maketitle

\section{\label{sec:level1}Introduction}

Research on primary cosmic rays (PCR) has been actively conducted for many decades. Recent studies~\cite{Thoudam_2016} indicate that a significant portion of PCR events in the energy range of 1 to 1000 PeV likely have an extragalactic origin. This discovery is crucial for developing theories on the transition of cosmic rays from galactic to extragalactic states and understanding their acceleration and propagation mechanisms.

The method of detecting Cherenkov light reflected off snow, proposed by A.E. Chudakov~\cite{chu74:VKKL74}, laid the foundation for the SPHERE series of experiments. Conducting experiments focused on the nuclear composition of cosmic rays is an important task in modern astrophysics. 

The SPHERE project team~\cite{Ant15a} is currently actively working on developing the next detector in the series, \mbox{SPHERE-3}. This detector~\cite{nucl2022} is intended to provide more detailed information about the mass of PCR in the specified energy range. The main research direction involves developing various detector geometries and verifying their consistency through multi-step modeling, including the use of software packages CORSIKA~\cite{corsika}, GEANT4~\cite{ALLISON2016186}, and specialized applications in FORTRAN.

\section{Detailing the Simulation Process of Cherenkov Images from Extensive Air Showers for the SPHERE-3 Detector}
\subsection{Design and key features of the detector project}

The SPHERE-3 detector incorporates an advanced Schmidt optical system with a lens corrector to eliminate spherical aberrations. Its structure, made of 20 mm diameter aluminum tubes, integrates key components of the optical system: the mirror, aperture, corrective lens, and a mosaic of silicon photomultipliers. The measuring equipment is placed in the shaded area of a truncated cone. 

The effective aperture area of the SPHERE-3 detector is expected to be at least 1 m\textsuperscript{2}, with an optical resolution of at least 2000 pixels. The planned field of view is at least ±20°. Final specifications will be determined after completing the optimization of the optical system parameters and modeling. 

A modified version of the Schmidt optical scheme is considered, featuring an aspherical mirror and a corrector plate. The telescope's entrance window is covered with an acrylic corrector plate, ranging from 5 to 30 mm in thickness and 1700 mm in diameter. Between the mirror and corrector lies the light-sensitive SiPM mosaic~\cite{ACERBI201916} with a sensitive part diameter of 660 mm and a total diameter of 680 mm, along with an electronics unit. Considering the shading caused by the mosaic and electronics unit, the effective aperture area is 1.9~m\textsuperscript{2}.

\subsection{The process of generating extensive air showers}

Traditionally, SPHERE telescopes have registered Cherenkov light from extensive air showers (EAS), reflected off the snow, by analyzing its spatiotemporal distribution. In experiments with the SPHERE-2 telescope, a signal preceding the reflected light was detected, interpreted as direct Cherenkov light passing through mirror holes. This led to the idea of registering direct light alongside with the reflected. The SPHERE-3 project considers various designs for direct light registration, including photomultiplier tube (PMT) mosaics and corrective lenses. Modeling EAS requires accounting for the distribution of Cherenkov light both on the snow surface and at different heights above it. Using the CORSIKA code, we optimize data processing, focusing on photons crucial for generating photoelectrons in sensors. The data are stored as multidimensional arrays for efficient analysis. Modeling parameters include primary energy, angles, particle types, and atmospheric and nuclear interaction models. The base sample size of EAS events is 100, sufficient for analyzing the main characteristics of the showers. These data are then used to generate larger samples of Cherenkov images by transposition.

For the generation of EAS event data in the SPHERE-3 project, the Lomonosov-2 supercomputer is employed. This powerful computational complex enables the processing of large volumes of information and the execution of complex calculations with a high degree of accuracy. Thanks to its high-performance capabilities, we can significantly reduce the time required for detailed analysis and modeling of the physical processes occurring during extensive air showers. The use of the Lomonosov-2 supercomputer accelerates the modeling process, enhances the accuracy of the data obtained, and efficiently processes the photon distributions for the subsequent creation of precise Cherenkov images.

\subsection{Turning Cherenkov light distribution on the snowed ground into the photon bunches on the aperture of the telescope}

The data obtained from CORSIKA simulations constitute massive samples containing comprehensive information about extensive air showers. For their effective analysis and interpretation, a specialized program written in FORTRAN is used. This stage of analysis is aimed at reconstructing a detailed image of the shower on the detector's diaphragm, including the identification of both primary and secondary parameters of the shower process.

Additionally, to increase the volume of statistical data without compromising its quality, an event cloning method is utilized. This process involves spatially shifting the shower axis relative to the telescope axis, allowing for the simulation of various observation scenarios based on multiple potential observation points.

An important part of this stage is also the integration of background events into the final data sets. This allows for simulating the operating conditions of the detector in a real environment, considering not only pure signals from showers but also various interference's and background effects. Such an approach ensures increased realism of the simulated scenarios and allows for more precise calibration and tuning of the detector before its practical application.

\subsection{Tracking photons through the telescope}

 Distributions of Cherenkov light, generated using FORTRAN software, are subsequently fed into a model developed with the Geant4 package. This tool allows for the simulation of photon trajectories from their inception at the detector's entrance window to their detection by the final elements of the detector mosaic. During this process, Geant4 considers multiple factors, including photon interactions with detector materials and their subsequent transformations.

Each registered particle is characterized by several parameters, recorded in the output file. These parameters include the pixel number where detection occurs, the precise registration time of the photoelectron in the detector mosaic, inherited from CORSIKA data and counted from the moment of primary particle interaction, and an informational key indicating whether the event is part of the main signal or background.

This detailed information about each particle plays a key role in assessing the accuracy and efficiency of the used models. Through additional data processing and analysis algorithms, we can conduct a thorough evaluation of both the quality of the modeling and potential improvements in the detector's design.

\subsection{Simulation parameters}

One of the primary objectives of the SPHERE-3 experiment is the detailed analysis and refinement of the mass composition of primary cosmic rays. To accurately assess the detector's capability in addressing these complex tasks, it is essential to generate a substantial volume of simulated EAS with a diverse range of parameters.

The following section provides an in-depth description of these simulation parameters:

\textbf{Type of Primary Particle}: The simulation currently encompasses six distinct types of particles: protons, alpha particles, and nuclei of nitrogen, aluminum, sulfur, and iron. This variety allows for a comprehensive understanding of the detector's response to different cosmic ray components.

\textbf{Energy of the Particle at Primary Interaction}: The simulation incorporates four discrete energy levels – 5, 10, 30, and 100 PeV. This range is crucial for examining the detector's sensitivity and efficiency across different cosmic ray energy spectra.

\textbf{Zenith Angle of the Shower Axis}: The simulation accounts for zenith angles ranging from 5 to 30 degrees, with an incremental step of 5 degrees. This parameter is vital for understanding the angular resolution and detection capabilities of the SPHERE-3 detector under various atmospheric conditions.

\textbf{Model of Hadron-Hadron Interaction}: Two models are employed, namely QGSJET01~\cite{KALMYKOV199717} and QGSJETII04~\cite{Ostapchenko_2014}. The use of multiple interaction models is essential for cross-validating the simulation results and ensuring the robustness of the findings.

\textbf{The Atmosphere Model}: 
In this research, we employ multiple atmospheric models from CORSIKA, specifically models : U.S. standard atmosphere~\cite{USStandardAtmosphere1976}, AT223 Central European atmosphere, AT511 Central European atmosphere, South pole atmosphere for March 31, 1997 (MSIS-90-E)
These models cater to various atmospheric scenarios impacting the behavior of EAS.

\subsection{Automating the generation and processing of Cherenkov shower images}

The simulation of Extensive Air Shower (EAS) detection by the SPHERE-3 telescope is a complex process, involving numerous steps and parameters. Each stage of this process demands meticulous verification of input data and careful calibration of operational parameters. These requirements significantly complicate the task of mass event generation. A schematic representation of the computational workflow is illustrated in the Fig. 1.

\begin{figure}[h]
    \centering
    \includegraphics[width=\linewidth]{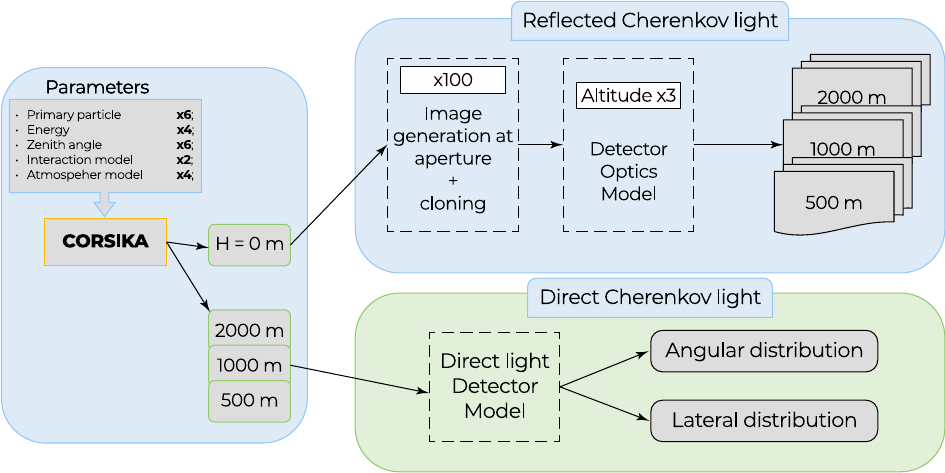}
    \caption{Workflow diagram.}
    \label{fig:pipeline}
\end{figure}

To address this challenge, a strategic decision was made to automate all stages using a Python co-routine. This approach has effectively minimized the risk of human error, thereby enhancing the accuracy and efficiency of the simulation process. The implementation of automation not only streamlines the workflow but also ensures a high level of precision in the simulation outcomes, mirroring the complexity and nuances of real-world data acquisition.

Furthermore, the application is being augmented with methods for further scaling the automation process. Future developments include the integration of image processing stages, along with the implementation of an electronics model that closely replicates the data handling observed in actual experimental scenarios. This advancement aims at providing a more holistic and realistic simulation environment, bridging the gap between theoretical models and practical experimental setups. The incorporation of these features will significantly enhance the capability of the SPHERE-3 simulation framework, offering a more robust and comprehensive tool for the study of cosmic ray interactions.

\section{Conclusion}

In this study, we presented a software suite for the multiple generations of Cherenkov light images on the detector mosaic of the SPHERE-3 telescope. The development of new detector geometries and the verification of established criteria for separating Extensive Air Showers (EAS) by primary mass require accumulating a substantial event dataset.

The automation of routine tasks and the integration of individual programs into a cohesive suite simplifies the statistical data collection process and reduces the impact of human error. Future enhancements of the software will incorporate and automate the processing stages of the obtained events, facilitating efficient validation of new detector geometries. This integration represents a significant step towards streamlining the analysis and enhancing the accuracy of cosmic ray detection and characterization.

\begin{acknowledgments}

This work was supported by a grant from the Russian Science Foundation No 23-72-00006, https://rscf.ru/project/23-72-00006/. The research is carried out using the equipment of the shared research facilities of HPC computing resources at Lomonosov Moscow State University~\cite{Lom-2}.

\end{acknowledgments}

\bibliography{refers}

\end{document}